\documentclass[twocolumn, tighten]{aastex61}
\usepackage{amsmath}	
\usepackage{amssymb}	
\usepackage{microtype}
\usepackage{booktabs}
\usepackage{multirow}
\usepackage{color}

\let\texdisplaystyle\displaystyle
\def\displaytotextstyle{\textstyle\let\displaystyle\texdisplaystyle}

\newcommand{\lya}{Ly$\alpha$}

\newcommand{\kms}{${\rm km\,s}^{-1}$}

\newenvironment{talign*}
 {\let\displaystyle\displaytotextstyle\csname align*\endcsname}
 {\endalign}

\submitjournal{ApJL}

\shorttitle{The gaseous halos of dwarf galaxies}
\shortauthors{Johnson et al.}

\begin{document}

\title{The extent of chemically enriched gas around star-forming dwarf galaxies}

\correspondingauthor{Sean D. Johnson}
\email{sdj@astro.princeton.edu}

\author[0000-0001-9487-8583]{Sean D. Johnson}
\altaffiliation{Hubble \& Carnegie-Princeton fellow}
\affil{Department of Astrophysical Sciences, 4 Ivy Lane, Princeton University, Princeton, NJ 08544, USA}
\affil{The Observatories of the Carnegie Institution for Science, 813 Santa Barbara Street, Pasadena, CA 91101, USA}

\author[0000-0001-8813-4182]{Hsiao-Wen Chen}
\affiliation{Department of Astronomy \& Astrophysics, The University of Chicago, 5640 S. Ellis Avenue, Chicago, IL 60637, USA}

\author[0000-0003-2083-5569]{John S. Mulchaey}
\affil{The Observatories of the Carnegie Institution for Science, 813 Santa Barbara Street, Pasadena, CA 91101, USA}

\author[0000-0002-0668-5560]{Joop Schaye}
\affil{Leiden Observatory, Leiden University, PO Box 9513, NL-2300 RA Leiden, the Netherlands}

\author[0000-0001-5892-6760]{Lorrie A. Straka}
\affil{Leiden Observatory, Leiden University, PO Box 9513, NL-2300 RA Leiden, the Netherlands}

\begin{abstract}

Supernova driven winds are often invoked
to remove chemically enriched gas from dwarf galaxies to match their low
observed metallicities.
In such shallow potential wells, outflows may produce
massive amounts of enriched halo gas (circum-galactic medium (CGM))
and pollute the intergalactic medium (IGM).
Here, we present a survey of the
CGM and IGM around $18$ star-forming field dwarfs
with stellar masses of $\log\,M_*/M_\odot\approx8-9$ at $z\approx0.2$.
Eight of these have CGM probed by quasar absorption spectra at projected
distances, $d$, less than the host virial radius, $R_{\rm h}$.
Ten are probed in the surrounding IGM at $d/R_{\rm h}=1-3$.
The absorption measurements
include neutral hydrogen, the dominant silicon ions
for diffuse cool gas ($T\sim10^4$ K; Si\,II, Si\,III, and Si\,IV),
moderately ionized carbon (C\,IV), and
highly ionized oxygen (O\,VI). Metal
absorption from the CGM of the dwarfs is
less common and $\approx4\times$ weaker compared to
massive star-forming galaxies though O\,VI absorption is still
common. None of the dwarfs probed at $d/R_{\rm h}=1-3$ have
definitive metal-line detections.
Combining the available silicon ions, we estimate that the
cool CGM of the dwarfs accounts for only $2-6\%$ of
the expected silicon budget from the yields of supernovae
associated with past star-formation.
The highly ionized O\,VI
accounts for $\approx8\%$ of the oxygen budget.
As O\,VI traces an ion with expected equilibrium ion fractions
of $\lesssim 0.2$, the highly ionized CGM may 
represent a significant metal reservoir even for dwarfs
not expected to maintain
gravitationally shock heated hot halos.

\end{abstract}

\keywords{galaxies: dwarf --- galaxies: halos --- intergalactic medium --- quasars: absorption lines}

\section{Introduction}
\label{section:introduction}

To match the observed low metallicities
\citep[e.g.][]{Lee:2006} and stellar masses
\citep[e.g.][]{McGaugh:2010} of low- and intermediate-mass galaxies,
models of galaxy evolution often invoke supernovae winds
to prevent gas cooling and drive metal enriched gas out
of star-forming regions \citep[for a review, see][]{Naab:2017}.
In the shallow gravitational potential wells of dwarf galaxies,
winds are expected to more easily reach large distances
resulting in a decreased efficiency of wind recycling,
a larger mass fraction of enriched halo gas
(the circum-galactic medium (CGM)),
and pollution of the surrounding intergalactic medium (IGM) with metals.

Alternatively, interaction-related turbulent mixing and heating may remove
enriched gas from galaxies without supernovae
winds \citep[e.g.][]{Tassis:2008}.
In this scenario, enriched material is not expected to escape from
the host halo resulting in enriched CGM
but pristine IGM around low-mass galaxies.
A census of metal enriched
CGM and IGM around dwarf galaxies therefore represents
a key test of galaxy evolution models.

While the IGM and CGM are typically too diffuse
to observe in emission with current facilities,
UV absorption spectra from the Cosmic Origins Spectrograph
\citep[COS;][]{Green:2012} on the {\it Hubble Space Telescope} (HST)
enable multi-phase surveys of the IGM/CGM at $z<0.4$
\citep[for reviews see][]{Chen:2016, Richter:2016review, Tumlinson:2017}.
For massive galaxies with stellar masses of $\log M_*/M_\odot \approx 10.5$,
the CGM detected in the UV and X-ray at projected distances, $d$, less than the
host halo virial radius, $R_{\rm h}$,
constitutes a significant metal reservoir
\citep[e.g.][]{Peeples:2014, Miller:2015}.
For less massive galaxies of $\log M_*/M_\odot < 9.5$,
stars and the observed interstellar medium (ISM)
only account for $\approx20\%$ of the expected metal yields
from supernovae associated with past star formation \citep[][]{Tumlinson:2017},
but surveys of the CGM and IGM around
dwarf galaxies are less extensive.

Recent studies of the CGM and IGM around intermediate-mass ($\log M_*/M_\odot \approx 9.5$) galaxies found little evidence for
metal absorption at projected distances of $d/R_{\rm h}\gtrsim0.7$
\citep[][]{Bordoloi:2014, Liang:2014}.
Moreover, the CGM of lower mass dwarf galaxies ($\log M_*/M_\odot < 9.0$) exhibit
lower C\,IV \citep[][]{Burchett:2016} and O\,VI \citep{Prochaska:2011} covering fractions
than massive galaxies.
Together, these observations suggest that galactic winds from dwarf galaxies may not pollute the CGM/IGM effectively or that their CGM is dominated by other ions.
Additional observations of low mass galaxies with measurements
for a wider range of metal ions are needed for a more comprehensive
view of the distribution and state of metal enriched gas around dwarf galaxies.

Here, we present a survey of diffuse absorbing gas around $18$
star-forming field dwarf galaxies
with a median stellar mass of $\log\,M_*/M_\odot=8.4$ at $z\approx0.1-0.3$.
Eight of these dwarf galaxies are probed by quasar spectra at projected
distances of $d/R_{\rm h}<1$. Ten are probed at
$d/R_{\rm h}=1-3$, enabling constraints
on chemical enrichment beyond the nominal halo radius.
The median stellar mass of our dwarf sample is an order of magnitude lower
than those of previous surveys with simultaneous measurements of absorption for the dominant silicon ions
for diffuse cool gas at $T\sim10^4$ K (Si\,II, Si\,III, and Si\,IV),
moderately ionized carbon (C\,IV) and
highly ionized oxygen (O\,VI) from COS spectra.
The field dwarf galaxy sample is based on
deep and highly complete redshift surveys
ensuring that the dwarfs are not satellites of massive galaxies.
Together, these data enable a comprehensive, controlled, and sensitive
study of enriched CGM and IGM around dwarf galaxies.

This Letter proceeds as follows. In Section \ref{section:data}
we describe the dwarf galaxy sample and
absorption spectroscopy. In Section \ref{section:CGM} we
characterize the CGM and IGM around the dwarfs.
In Section \ref{section:discussion} we discuss the implications of the survey.

Throughout the Letter we adopt a flat $\Lambda$ cosmology
with $\Omega_{\rm m}=0.3$, $\Omega_\Lambda=0.7$,
and $H_0=70\,{\rm km\,s^{-1}\,Mpc^{-1}}$. We define a galaxy with luminosity
of $L=L_*$ as having an $r$-band absolute magnitude of $M_r=-21.5$
based on the luminosity function
from \cite{Loveday:2012}. All of the quoted magnitudes are in the $AB$ system.

\section{Sample and data description}
\label{section:data}

\begin{longrotatetable}
\begin{deluxetable*}{lcccccrrcccccc}
\tablenum{1}
\tablecaption{Summary of dwarf galaxy properties and associated absorption}
\label{table:survey}
\tablehead{
\colhead{} & 
\colhead{} &
\colhead{} & 
\colhead{} &
\colhead{} & 
\colhead{} &
\colhead{} &
\colhead{} &
\multicolumn{5}{c}{$W_{\rm r}$}
\\
\cline{9-14}
\colhead{} & 
\colhead{R.A.} &
\colhead{Dec.} & 
\colhead{} &
\colhead{$M_r$} & 
\colhead{} &
\colhead{$d$} &
\colhead{$R_{\rm h}$} &
\colhead{${\rm H\,I\,\,1215}$} &
\colhead{${\rm Si\,II\,\,1260}$} &
\colhead{${\rm Si\,III\,\,1206}$} &
\colhead{${\rm Si\,IV\,\,1393}$} &
\colhead{${\rm C\,IV\,\,1548}$} &
\colhead{${\rm O\,VI\,\,1031}$}
\\
\colhead{ID} & 
\colhead{(J2000)} &
\colhead{(J2000)} & 
\colhead{$z_{\rm gal}$} &
\colhead{($AB$)} & 
\colhead{$\log\,M_*/M_\odot$} &
\colhead{(kpc)} &
\colhead{(kpc)} &
\colhead{(\AA)} &
\colhead{(\AA)} &
\colhead{(\AA)} &
\colhead{(\AA)} &
\colhead{(\AA)} &
\colhead{(\AA)}
}

\startdata
D1	& 06:35:44.6 & $-$75:16:18 & $0.1229$  & $-15.9$ & $7.9$  & $16$  & $70$   & $0.69 \pm 0.01$ & $<0.02$     & $0.13 \pm 0.01$ & $$ & $0.13 \pm 0.01$ & $0.12 \pm 0.01$ \\
D2	& 06:35:45.7 & $-$75:16:23 & $0.1613$  & $-16.2$ & $8.1$  & $21$  & $80$   & $0.77 \pm 0.10$ & $<0.01$     & $0.07 \pm 0.01$ & $<0.02$ & $$              & $0.13 \pm 0.01$ \\
D3	& 02:35:07.7 & $-$04:02:14 & $0.2960$  & $-17.7$ & $8.8$  & $48$  & $110$  & $0.41 \pm 0.01$ & $<0.02$     & $<0.02$         & $<0.10$ & $<0.09$         & $<0.02$         \\
D4  & 10:04:02.3 & $+$28:55:12 & $0.1380$  & $-17.2$ & $8.2$  & $56$  & $80$   & $0.69 \pm 0.01$ & $<0.01$     & $$              & $<0.03$ & $<0.04$         & $0.11 \pm 0.01$ \\
D5 	& 06:35:44.3 & $-$75:15:55 & $0.1436$  & $-17.4$ & $8.6$  & $57$  & $100$  & $1.14 \pm 0.07$ & $<0.01$     & $$              & $<0.02$ & $$              & $<0.01$         \\
D6	& 14:37:49.6 & $-$01:47:03 & $0.1839$  & $-19.0$ & $9.2$  & $63$  & $130$  & $0.54 \pm 0.01$ & $<0.01$     & $<0.01$         & $<0.01$ & $<0.07$         & $0.13 \pm 0.01$ \\
D7	& 04:07:49.3 & $-$12:12:16 & $0.0923$  & $-16.1$ & $8.3$  & $72$  & $90$   & $0.47 \pm 0.01$ & $<0.01$     & $$              & $<0.02$ & $<0.01$         & $0.08 \pm 0.01$ \\
D8  & 15:47:45.4 & $+$20:51:41 & $0.0949$  & $-17.3$ & $8.5$  & $79$  & $100$  & $<0.01$         & $<0.02$     & $<0.01$         & $$      & $<0.03$         & $<0.02$         \\
D9	& 15:24:23.3 & $+$09:58:58 & $0.1391$  & $-14.9$ & $7.7$  & $84$  & $60$   & $0.06 \pm 0.01$ & $<0.02$     & $<0.01$         & $<0.04$ & $<0.02$         & $0.05 \pm 0.01$ \\
D10 & 15:55:47.7 & $+$11:11:20 & $0.1234$  & $-16.7$ & $8.0$  & $155$ & $80$   & $0.07 \pm 0.01$ & $<0.01$     & $<0.01$         & $<0.02$ & $<0.02$         & $<0.01$         \\
D11	& 10:05:32.2 & $+$01:33:45 & $0.1245$  & $-16.9$ & $8.1$  & $168$ & $80$   & $0.06 \pm 0.01$ & $<0.03$     & $<0.02$         & $<0.02$ & $<0.03$         & $<0.03$         \\
D12	& 15:24:22.9 & $+$09:59:07 & $0.2402$  & $-17.7$ & $8.9$  & $169$ & $110$  & $0.23 \pm 0.01$ & $<0.01$     & $$              & $<0.02$ & $<0.07$         & $<0.02$         \\
D13	& 14:37:44.5 & $-$01:48:10 & $0.1161$  & $-18.7$ & $9.0$  & $173$ & $120$  & $0.03 \pm 0.01$ & $<0.01$     & $<0.01$         & $<0.02$ & $<0.02$         & $<0.03$         \\
D14	& 02:35:07.5 & $-$04:03:10 & $0.1815$  & $-17.2$ & $8.2$  & $199$ & $80$   & $0.23 \pm 0.01$ & $<0.01$     & $<0.01$         & $<0.02$ & $<0.10$         & $<0.02$         \\
D15	& 15:24:26.9 & $+$09:57:31 & $0.1981$  & $-16.9$ & $8.2$  & $222$ & $80$   & $0.10 \pm 0.02$ & $<0.01$     & $$              & $<0.04$ & $<0.12$         & $<0.01$         \\
D16	& 02:28:20.7 & $-$40:56:49 & $0.1992$  & $-18.3$ & $9.2$  & $224$ & $130$  & $0.35 \pm 0.01$ & $<0.01$     & $<0.01$         & $<0.03$ & $$              & $<0.01$         \\ 
D17	& 02:28:11.3 & $-$40:56:36 & $0.2804$  & $-17.8$ & $8.4$  & $244$ & $90$   & $0.18 \pm 0.02$ & $<0.01$     & $<0.01$         & $<0.03$ & $$              & $<0.01$         \\ 
D18	& 14:37:49.1 & $-$01:48:10 & $0.2716$  & $-17.1$ & $8.8$  & $256$ & $120$  & $0.28 \pm 0.01$ & $<0.01$     & $<0.01$         & $<0.02$ & $<0.07$         & $<0.01$         \\
\enddata
\end{deluxetable*}
\end{longrotatetable}

\subsection{Field dwarf galaxy sample}

\begin{figure}[t!]
\centering
	\includegraphics[scale=0.43]{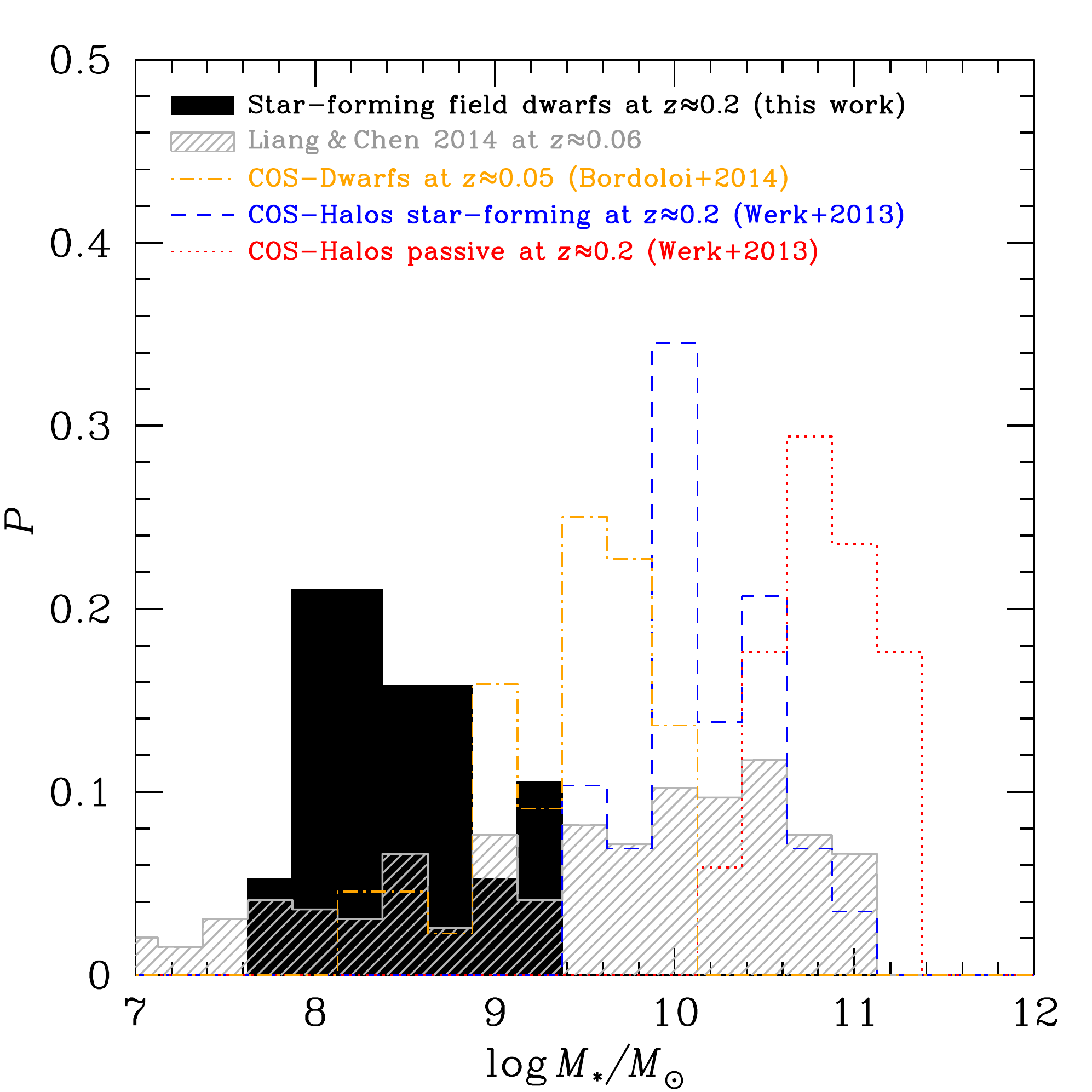}
	\caption{Stellar mass distribution of the star-forming field
	         dwarf galaxies. For comparison, the
	         star-forming and passive galaxies from the COS-Halos survey
	         and intermediate mass samples from COS-Dwarfs and \cite{Liang:2014}
	         are also shown.}
	\label{figure:logMstar}
\end{figure}
The dwarf galaxy sample is drawn from our
absorption-blind spectroscopic surveys in the fields
of quasars
with high $S/N$ COS absorption spectra.
The redshift surveys were conducted with
the Magellan telescopes targeting
galaxies of $r<23.5$ mag at angular
distances of $\Delta \theta < 10'$ from the quasars.
At redshifts
of $z=0.1-0.3$, the magnitude limit corresponds to
$L/L_*=0.0025-0.025$.
The details of the Magellan survey
are published in \cite{Chen:2009},
and \cite{Johnson:2015a}.
In summary the redshift errors are
$\Delta v = 30-60$ \kms, stellar masses assume a \cite{Chabrier:2003}
initial mass function and $g-r$ color dependent
mass-to-$r$-band light ratio, halo masses assume
the stellar-to-halo mass function from \cite{Kravtsov:2014},
and virial radii are defined as in \cite{Bryan:1998}.

To construct a field dwarf galaxy sample from our Magellan survey, we identified
galaxies at $z=0.09-0.3$, of $L<0.1\,L_*$, and
at projected distances of $d/R_{\rm h}<3$ from the quasar sightline.
The redshift cut ensures coverage of the desired absorption
features
and that the redshift surveys are sufficient to
differentiate field dwarfs from satellites.
The luminosity cut limits the sample to low-mass galaxies.
The projected distance cut is chosen to be larger than $R_{\rm h}$
to examine enriched gas that may have escaped
from the dwarfs.
To ensure that the galaxy sample
does not include satellites of massive galaxies, we removed
galaxies that are within a projected distance of $500$ kpc and
radial velocity of $\pm300$ \kms\, of a galaxy with $L>0.1\,L_*$
(40\% sample reduction).
We supplement our Magellan redshift survey
with two field dwarf galaxies from \cite{Chen:1998}.

The properties of the resulting sample of $18$ dwarf galaxies
are summarized in
Table \ref{table:survey}.
The sample  spans a stellar
mass range of $\log M_*/M_\odot = 7.7-9.2$ with a median of
$\log M_*/M_\odot = 8.4$ (see
Figure \ref{figure:logMstar}).
All of the dwarfs galaxies have blue colors and multiple
detected emission lines consistent with
the low quiescent fraction of dwarfs in the
field \citep[e.g.][]{Peng:2010}.

\subsection{COS absorption spectroscopy}

To study the CGM and IGM around the dwarf galaxies,
we searched for H\,I and metal ion
absorption in the COS quasar spectra within $\pm 300$ \kms\
of the galaxy systemic redshifts.
Increasing the search window to $\pm 600$ \kms\,
does not change the results for metal ions.
Moreover, the velocity window is $>2\times$
the estimated escape velocity in all cases
assuming an Nararrow-Frenk-White (NFW) profile \citep{Navarro:1996}
with a concentration of $10$ \citep[][]{Correa:2015}.

For galaxies with detected absorption, we measured the rest-frame
equivalent widths, $W_{\rm r}$,
of H\,I \lya, Si\,II $\lambda 1260$, Si\,III $\lambda 1206$,
Si\,IV $\lambda 1393$, C\,IV $\lambda 1548$, and O\,VI $\lambda 1031$
after local continuum normalization.
For non-detections,
we report 2-$\sigma$ upper limits
integrated over $90$ \kms\
(the median FWHM of O\,VI
absorption from \cite{Johnson:2015a}).
When transitions are contaminated by systems
at distinct redshifts, we placed limits based on
other available transitions for the same ion
accounting for oscillator strength and wavelength differences.
The absorption equivalent widths and limits are
in Table \ref{table:survey}. For D7, we adopt absorption measurements from \cite{Prochaska:2004}.

\section{Results}
\label{section:CGM}

\begin{figure*}
	\centering
	\includegraphics[scale=0.65]{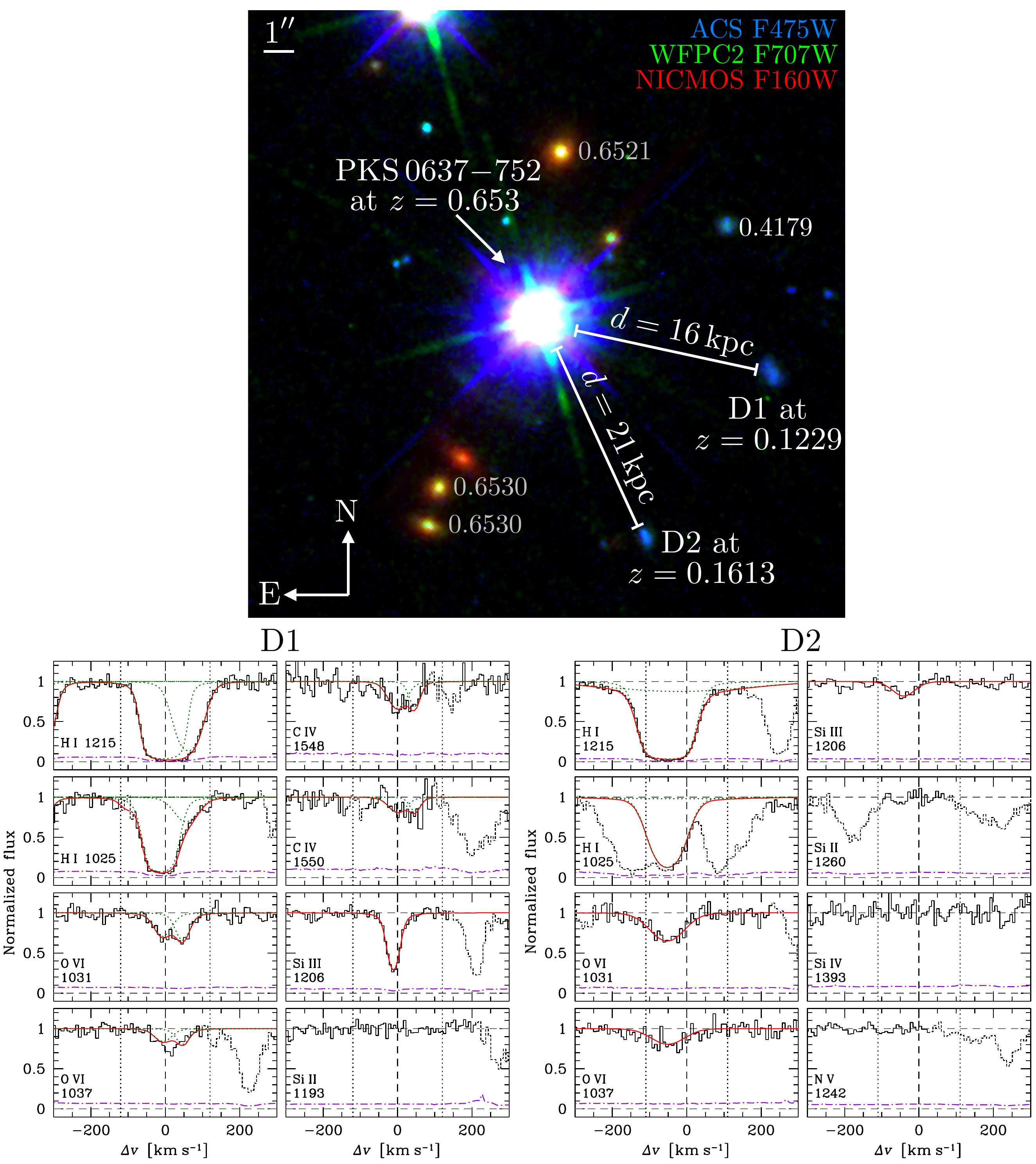}
	\caption{Examples of the star-forming field dwarf galaxies D1 and D2 near the sightline toward PKS\,0637$-$752. The {\it top} panel shows a color composite HST image (20'' on a side). Galaxies foreground to the quasar and those near or background to the quasar are labeled with their redshifts in white and grey respectively. The panels on the {\it bottom left} and {\it right} show the continuum normalized COS absorption spectra for H\,I and metal ion transitions with zero velocity set by the redshifts of D1 and D2. The COS spectrum, error array, best-fit Voigt model, and components are shown in black histogram, purple dash-dotted line, red solid line, and green dashed line, respectively. The estimated escape velocities for D1 and D2 ($\approx120$ \kms) are marked by vertical dotted lines. Spectral regions contaminated by absorption from systems at distinct redshifts are dotted out.}
	\label{figure:D1_D2}
\end{figure*}

With the survey data
described in Section \ref{section:data}, we characterize the
CGM and IGM around star-forming field dwarf galaxies
first by highlighting the two probed
at the smallest projected distances in the sample
in Figure \ref{figure:D1_D2}.
These two dwarfs, D1 and D2, are at $z=0.1229$ and $0.1613$,
respectively, and have estimated stellar masses
of $\log\,M_*/M_\odot = 7.9$ and $8.1$, respectively.
Non-detection of [N\,II] $\lambda 6585$ lead to
2$\sigma$ upper limits on their ISM metallicities of
$12+\log {\rm O/H}<8.4$
\citep[following][]{Pettini:2004}
D1 and D2 are probed
by the quasar sightline at $d=16$ and $21$ kpc
or $d/R_{\rm h}=0.2$ and $0.3$ respectively.
For both D1 and D2, the COS spectra reveal strong H\,I
absorption as well as
Si\,III and O\,VI absorption. C\,IV absorption is detected for D1 but
not covered by COS for D2.
Si\,IV absorption is not detected for D2
but is not covered for D1.
Neither dwarf is detected in Si\,II.
Apart from a wing on the O\,VI line observed for D2,
the metal absorption kinematics
are consistent with bound motion in the host halos
of D1 and D2, which have estimated escape velocities of $\approx120$
\kms.

\startlongtable
\begin{deluxetable*}{lccccccccc}
\tablenum{2}
\tablecaption{Voigt profile fitting measurements for D1 and D2}
\label{table:voigt}
\tablehead{
\colhead{} &
\colhead{} &
\colhead{} &
\colhead{$\Delta v$} &
\colhead{$b$} &
\colhead{} &
\colhead{} &
\colhead{} \\
\colhead{Quasar field} &
\colhead{ID} &
\colhead{Species} &
\colhead{(\kms)} &
\colhead{(\kms)} &
\colhead{$\log N/{\rm cm}^{-2}$} &
\colhead{$\log N_{\rm tot}/{\rm cm}^{-2}$} &
\colhead{Method}
}

\startdata
PKS\,0637$-$752    & D1  & H\,I    & $-12\pm3$    & $27^{+2}_{-6}$    & $15.7^{+0.6}_{-0.2}$   & $15.7^{+0.6}_{-0.2}$    & MCMC          \\
                   &     & H\,I    & $48\pm7$     & $41^{+12}_{-3}$   & $14.3^{+0.3}_{-0.1}$   & $$                      & MCMC          \\
                   &     & Si\,II  &              &                   & $<12.15$               & $<12.15$                & n/a           \\
                   &     & Si\,III & $-16\pm1$    & $15\pm1$          & $13.14\pm0.03$         & $13.14 \pm 0.03$        & VPFIT         \\
                   &     & C\,IV   & $-7\pm12$    & $30\pm15$         & $13.53\pm0.19$         & $13.73\pm0.04$          & VPFIT         \\
                   &     & C\,IV   & $39\pm8$     & $16\pm11$         & $13.31\pm0.30$         & $$                      & VPFIT         \\
                   &     & O\,VI   &  $-12\pm12$  & $26\pm16$         & $13.79\pm0.19$         & $14.10 \pm 0.03$        & VPFIT         \\
                   &     & O\,VI   &   $41\pm8$   & $20\pm11$         & $13.80\pm0.18$         & $$                      & VPFIT         \\
PKS\,0637$-$752    & D2  & H\,I    & $-53\pm1$    & $46\pm1$          & $15.04\pm0.02$         & $15.06 \pm 0.02$        & VPFIT         \\
                   &     & H\,I    & $-23\pm30$   & $250\pm50$        & $13.66\pm0.05$         & $$                      & VPFIT         \\
                   &     & Si\,II  &              &                   & $<11.8$                & $<11.8$                 & n/a           \\
                   &     & Si\,III & $-41\pm4$    & $35\pm6$          & $12.48\pm0.05$         & $12.48 \pm 0.05$        & VPFIT         \\
                   &     & Si\,IV  &              &                   & $<12.3$                & $<12.3$                 & n/a           \\
                   &     & O\,VI   & $-53\pm3$    & $62\pm5$          & $14.17\pm0.02$         & $14.17 \pm 0.02$        & VPFIT         \\
\enddata
\end{deluxetable*}

To measure the column densities
for the absorption associated with D1 and D2, we performed Voigt profile
fitting with VPFIT \citep[][]{Carswell:1987}. For D1, the available Lyman series 
transitions are saturated so we estimate the column density range
allowed by the data using the EMCEE Markov chain Monte Carlo package
\citep[][]{Foreman-Mackey:2013}.
The Voigt profile measurements
are reported in Table \ref{table:voigt}. For non-detections,
we converted the equivalent width limits to column densities
assuming the linear portion of the curve of growth.

The measured H\,I column densities for D1 and D2 of
$\log N({\rm H\,I})/{\rm cm^{-2}} = 15.7^{+0.6}_{-0.2}$
and $15.06\pm0.02$ are lower than
those of the Lyman limit systems 
($\log N({\rm H\,I})/{\rm cm^{-2}} >17.2$)
often observed for star-forming galaxies of $\log M_*/M_\odot \approx 10.5$
at similar $d/R_{\rm h}$ \citep[e.g.][]{Prochaska:2017}. The
metal ion column densities observed for
D1 and D2 are lower as well with 
$\log N({\rm Si\,III})/{\rm cm^{-2}} = 13.14, 12.48$ and
$\log N({\rm O\,VI})/{\rm cm^{-2}} = 14.10, 14.17$,
respectively, compared to 
$\log N({\rm Si\,III})/{\rm cm^{-2}} \gtrsim 13.5$ \citep[][]{Werk:2013}
and $\log N({\rm O\,VI})/{\rm cm^{-2}} = 14.7$ \citep[][]{Tumlinson:2011}
for massive star-forming galaxies.

\begin{figure*}
	\centering
	\includegraphics[scale=0.8]{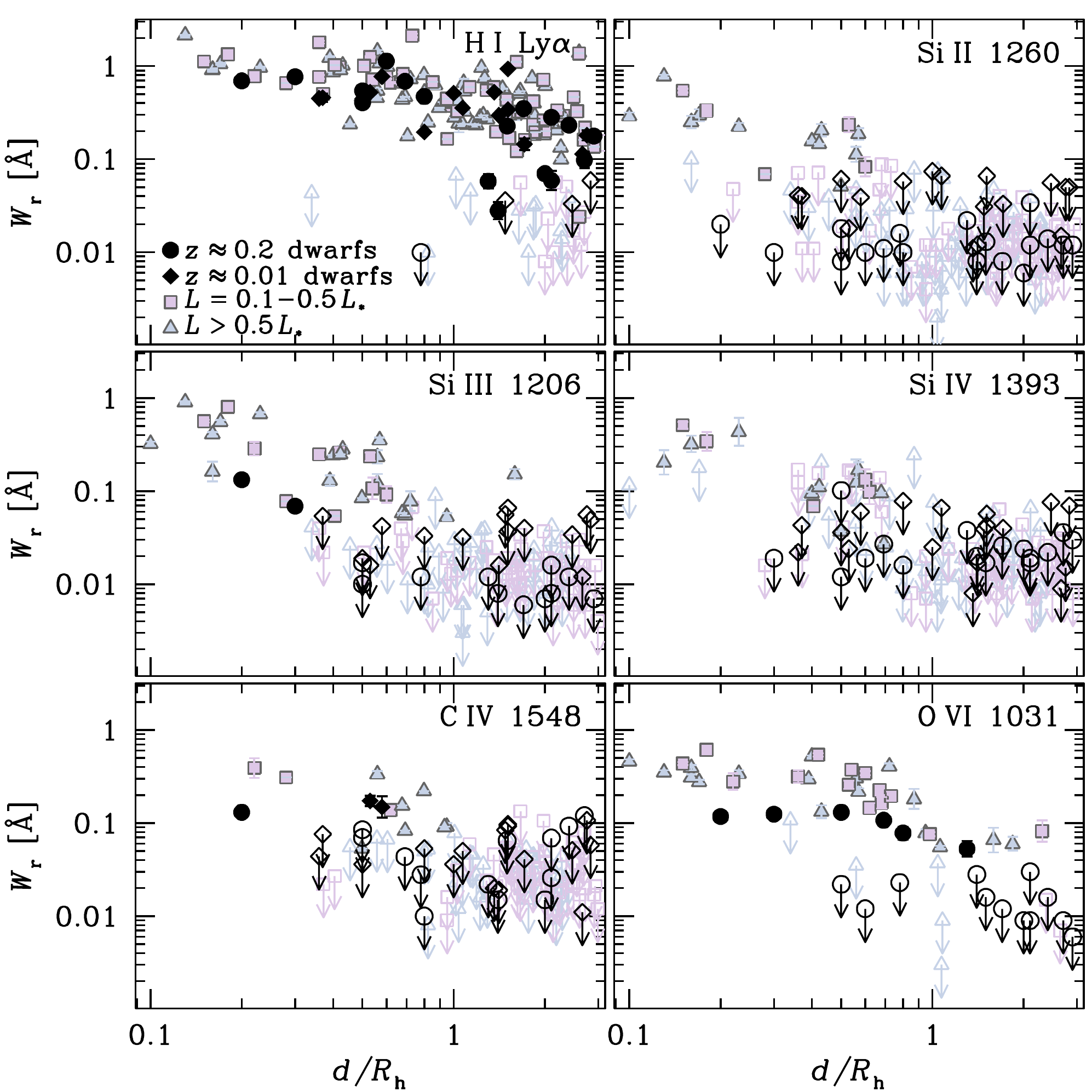}
	\caption{Rest-frame absorption equivalent width, $W_{\rm r}$, versus virial radius normalized projected distance, $d/R_{\rm h}$, for H\,I \lya, Si\,II $\lambda 1260$, Si\,III $\lambda1206$, Si\,IV $\lambda1393$, C\,IV $\lambda 1548$, and O\,VI $\lambda 1031$. The star-forming field dwarf galaxies at $z\approx0.2$ from this work are displayed as black circles and those of
$\log M_*/M_\odot < 9$ from the low redshift surveys of \cite{Liang:2014} and \cite{Bordoloi:2014} are displayed as black diamonds. For comparison, star-forming galaxies of $0.1\,L_*<L<0.5\,L_*$ and $L>0.5\,L_*$ from \protect \cite{Werk:2013} and \protect\cite{Johnson:2015a} are shown in faded violet squares and faded blue triangles respectively. Upper limits are $2\sigma$ and shown as open symbols with downward arrows.}
	\label{figure:W_vs_d_Rh}
\end{figure*}

To investigate whether the modest absorption levels
observed for D1 and D2 are typical of field dwarfs,
we plot the measured rest-frame equivalent widths, $W_{\rm r}$, of H\,I \lya,
Si\,II $\lambda 1260$, Si\,III $\lambda 1206$, Si\,IV $\lambda 1393$,
C\,IV $\lambda 1548$ and O\,VI $\lambda 1031$ absorption as a function
of $d/R_{\rm h}$ for the 18 dwarf galaxies in Figure \ref{figure:W_vs_d_Rh}.
To increase the available sample for all but O\,VI, we also
include star-forming field dwarfs of $\log\,M_*/M_\odot < 9$
from the low-redshift surveys
of \cite{Bordoloi:2014} and \cite{Liang:2014}
after ensuring similar environmental criteria, stellar mass estimates, and
absorption techniques.
The stellar mass cut was chosen to
a match the median stellar mass of our field dwarfs.
Massive star-forming galaxies
from \cite{Werk:2013}, \cite{Tumlinson:2013}, \cite{Liang:2014}, \cite{Johnson:2015a},
are shown for comparison.

Both H\,I and metal ion absorption
around the dwarf galaxies is anti-correlated with $d/R_{\rm h}$.
Despite the comparatively low H\,I column densities observed for D1 and D2, the \lya\,
equivalent widths of the dwarf sample overlap
with those of more massive galaxies (see the top left panel of Figure \ref{figure:W_vs_d_Rh}). This suggests that bulk motion
contributes to the equivalent widths of \lya\ observed
around galaxies. Confirmation will require larger samples
observed in the full Lyman series.

For metal ions on the other hand, the full sample confirms that the
star-forming field dwarf galaxies exhibit modest absorption levels
and lower covering fractions compared to more massive galaxies.
To better quantify this trend, we estimate
the covering fractions for
systems of $W_{\rm r}>0.1$ \AA\ for the silicon ions,
C\,IV, and O\,VI both in the CGM at $d<R_{\rm h}$ and IGM
at $d/R_{\rm h}=1-3$.

First we consider silicon ions.
None of the dwarf galaxies are detected
in Si\,II absorption with limits
of $W_{\rm r}({\rm Si\,II\, 1260})<0.01-0.1$ \AA\
(top right panel of Figure \ref{figure:W_vs_d_Rh}) leading
to 95\% upper limits on the covering fraction of
$\kappa_{\rm Si\,II}<0.17$ ($0.13$)
at $d/R_{\rm h}\le1$ ($1<d/R_{\rm h}\le 3$).
Only D1 exhibits Si\,III $\lambda 1206$ with $W_{\rm r}>0.1$ \AA\ 
with similar limits
(middle left panel of Figure \ref{figure:W_vs_d_Rh}) resulting in
covering fraction estimates of $\kappa_{\rm Si\,III}=0.10^{+0.17}_{-0.03}$
($<0.14$) at $d/R_{\rm h}\le1$ ($1<d/R_{\rm h}\le 3$).
Finally, none of the dwarf galaxies are detected
in Si\,IV with similar limits
(middle right panel of Figure \ref{figure:W_vs_d_Rh}) resulting in
covering fraction limits of $\kappa_{\rm Si\,IV}<0.19$ (0.13)
at $d/R_{\rm h}\le1$ ($1<d/R_{\rm h}\le 3$).

For moderately ionized carbon,
only D1 and two low-redshift dwarf galaxies are detected in C\,IV
with $W_{\rm r}>0.1$ \AA.
All three of these are probed at $d/R_{\rm h}<0.6$.
The COS spectra place limits of $W_{\rm r}({\rm C\,IV\, 1548})<0.01-0.1$ \AA\
for the other dwarfs (bottom left panel of Figure \ref{figure:W_vs_d_Rh}) leading
to covering fraction estimates of $\kappa_{\rm C\,IV}=0.23^{+0.15}_{-0.08}$
($<0.14$) at $d/R_{\rm h}\le1$ ($1<d/R_{\rm h}\le 3$),
consistent
with low C\,IV covering fractions observed around dwarf
galaxies by \cite{Burchett:2016}.

Considering highly ionized oxygen, D1 and D2
along two other dwarf galaxies are 
detected in O\,VI with $W_{\rm r}>0.1$ \AA\
and two show weaker O\,VI absorption
(bottom right panel of Figure \ref{figure:W_vs_d_Rh}).
Of the O\,VI detected dwarfs, only D9 is probed at $d/R_{\rm h}>1$,
but, the identification of the weak absorption as O\,VI is tentative due
to insufficient S/N to detect the weaker $\lambda 1037$ doublet member.
For the twelve dwarf galaxies without O\,VI detections, the COS
spectra place upper limits of $W_{\rm r}({\rm O\,VI\, 1031})<0.03$ \AA\ or better.
Based on this, we estimate an O\,VI covering fraction of $\kappa_{\rm O\,VI}=0.50^{+0.16}_{-0.16}$ ($\le 0.24$) at $d/R_{\rm h}\le1$ ($1<d/R_{\rm h}\le 3$) for
systems of $W_{\rm r}>0.1$ \AA,
consistent with non-detections of such
systems among the three late-type field dwarfs from
\cite{Prochaska:2011} which are probed
at $d/R_{\rm h}=0.5, 1.3$, and $1.4$.

\section{Discussion}
\label{section:discussion}
In a sample of star-forming field dwarf galaxies
with a median stellar mass of $\log M_*/M_\odot = 8.4$
probed at $d/R_{\rm h}<3$ by
absorption spectra,
we find that metal ion absorption systems
from a wide range of silicon ions
(Si\,II, Si\,III, Si\,IV)
and moderately ionized carbon (C\,IV)
are less common and weaker than observed
in the CGM of more massive star-forming galaxies.
Highly ionized oxygen absorption from O\,VI is
common in the CGM of the dwarf galaxies but with lower covering
fractions and absorption levels as well.
For example, the two dwarf galaxies probed at $d/R_{\rm h}=0.2-0.3$
exhibit Si\,III, C\,IV, and O\,VI column densities that are
$\approx 0.6$ dex lower than observed
around more massive star-forming galaxies.
None of the dwarf galaxies probed at $d/R_{\rm h}=1-3$
are definitively detected in metal ion absorption.
These observations stand in contrast to analytic expectations
and simulations 
\citep[e.g.][]{Shen:2014, Christensen:2016, Oppenheimer:2016, Muratov:2017, Wang:2017}
that predict more extended chemical enrichment and/or lower ionization
states around dwarf galaxies.

CGM studies often scale projected distances
by the estimated host virial radii
to account for the self-similar nature of dark matter halo
density profiles and approximate the observed
mass-size scaling of the CGM \citep[e.g.][]{Chen:2010a}.
However, NFW profiles are not self-similar in 
projected density.
An NFW profile with $c=10$ and $\log M_{\rm h}/M_\odot=10.8$
(corresponding to $\log M_{\rm *}/M_\odot=8.4$)
has projected densities that are $\approx0.3$ dex
lower than that a halo of $\log M_{\rm h}/M_\odot=12.0$
at fixed $d/R_{\rm h}$.
As this is smaller than the measured differences,
mass$-$size scaling of the CGM alone
cannot explain the decreased metal absorption
around the dwarfs.

Stars and the ISM of dwarf galaxies
account for $\approx20\%$ of the expected supernovae metal yields
associated with past star-formation
\citep[][]{Tumlinson:2017}, and
the CGM represents a potential reservoir
for the remaining metals.
We estimate metal mass of the CGM around dwarf galaxies
in each ion, $M_{\rm ion}$,
as
$M_{\rm ion}\approx \pi R_{\rm h}^2 m_{\rm ion} \kappa_{\rm ion}\langle N_{\rm ion} \rangle$ where $m_{\rm ion}$ is the atomic mass
and $\langle N_{\rm ion} \rangle$ is the mean column density among
the detections.
Adopting a virial radius of $R_{\rm h}=90$ kpc corresponding
to a dwarf of $\log M_*/M_\odot=8.4$, mean column densities of D1 and D2,
and observed covering fractions, we find
ion masses of $M_{\rm ion}\approx 10^4, 3\times10^4,$ and $3\times10^5 M_\odot$
for Si\,III, C\,IV, and O\,VI respectively.
By assuming unity covering fractions and the limits for D2,
we place limits of $M_{\rm ion}<5\times10^3$
and $10^4 M_\odot$ for Si\,II and Si\,IV.
Together, Si\,II, Si\,III, and Si\,IV are the dominant
silicon ions for diffuse gas
at $T=10^4-10^5 $ K \citep[][]{Oppenheimer:2013} so we estimate the total
silicon mass in the cool CGM phase of $1-2.5\times10^4 M_\odot$.

To place these metal masses in context, we compare them to
the metal budget from the yields of supernovae
associated with the formation of
$\log M_*/M_\odot=8.4$ worth of stars.
Adopting an IMF normalized supernovae oxygen yield of $0.015$ $M_\odot$
per solar mass of star formation
\citep[see][]{Peeples:2014} and
solar relative abundance \citep[][]{Asplund:2009},
we find that the cool CGM
represents just $2-6\%$ of the silicon budget of
the dwarf galaxies. The inferred ion
masses from C\,IV and O\,VI absorption represent
$\approx2$\% and $\approx8$\% of the expected carbon and oxygen
budgets. O\,VI traces an ion with expected
ion fractions of $\lesssim0.2$
under photo- or collisional ionization equilibrium \citep[][]{Oppenheimer:2013}
so this highly ionized CGM phase
may account for $\approx40\%$ of the metal budget of the dwarfs
though we caution that higher ion fractions are
possible \citep[e.g.][]{Vasiliev:2015}.
Dwarf galaxies are not expected to maintain
gravitationally shock heated hot halos \citep[e.g.][]{Correa:2017} suggesting
that the O\,VI arises from photoionized gas or other
shock/heating sources \citep[e.g.][]{Cen:2011}.

In summary, our observations of modest metal ion absorption
from Si\,II, Si\,III, Si\,IV, C\,IV, and O\,VI
in the CGM and IGM around star-forming field dwarf galaxies suggest that:
(1) the CGM of field dwarfs is dominated by unexpectedly high ionization states,
(2) galactic outflows are unexpectedly ineffective at enriching the cool-warm
IGM surrounding low-mass galaxies at $z<0.3$, and/or (3) enriched outflows effectively mix with the IGM over time to reach lower metallicities than are currently observable.

Dwarf galaxies are abundant and may contribute non-negligibly to
the low redshift quasar absorption populations
despite their modest CGM absorption.
We estimate the number of metal absorption systems
per unit redshift from field dwarf galaxies
of $\log M_*/M_\odot=8-9$ as
$\frac{dN}{dz}=\frac{c}{H_0}\Phi \sigma_{\rm ion}$,
where $\Phi$ is the number density
and $\sigma_{\rm ion}$ is the effective cross-section
from the virial radius and observed covering fractions.
Adopting $\Phi=0.017\,\,{\rm Mpc^{-3}\,dex^{-1}}$
based on the stellar mass function from \cite{Baldry:2008}
and discounting by the fraction
dwarfs that are satellites \citep[30\%;][]{Zheng:2007},
we find $\frac{dN}{dz}\approx0.1$ and $\approx0.7$ for Si\,III and O\,VI
systems with $W_{\rm r}\gtrsim0.1$ \AA.
Star-forming field dwarfs therefore account for $\approx5\%$ and $\approx20\%$ of such intervening Si\,III and O\,VI systems at low redshift
\citep[][]{Thom:2008, Tripp:2008, Danforth:2016}.

\section*{Acknowledgements}
We are grateful to Dave Bowen, Renyue Cen, Johnny Greco, Jenny Greene, Evan Schneider,
and Fakhri Zahedy for comments on the draft.
SDJ is supported by a NASA Hubble Fellowship (HST-HF2-51375.001-A).
HWC is partially supported by HST-GO-14145.01.A.
This research used data from the Magellan Telescopes and the NASA/ESA Hubble Space Telescope archive (NAS-5-2655: 6619, 10541,11508, 11520,11541,11692, 11741, 12025, 12038, 12264, 13024, 13398),
the NASA/IPAC Extragalactic Database,
and NASA Astrophysics Data System.

\vspace{5mm}
\facilities{HST, Magellan}

\bibliographystyle{aasjournal}
\bibliography{manuscript}

\end{document}